\documentclass[12pt]{article} 
\usepackage{cite,epsfig,axocolor}
 \def\baselinestretch{1.3}
\newcommand{\ba}{\begin{array}}
\newcommand{\ea}{\end{array}}
\newcommand{\bd}{\begin{displaymath}}
\newcommand{\ed}{\end{displaymath}}
\newcommand{\be}{\begin{equation}}
\newcommand{\ee}{\end{equation}}
\newcommand{\bea}{\begin{eqnarray}}
\newcommand{\eea}{\end{eqnarray}}

\def\dis{\displaystyle}
\def\barr{\begin{array}}
\def\earr{\end{array}} 
                               
                              \def\tev{\: \rm TeV}

\newcommand{\beqn}{\begin{eqnarray}}
\newcommand{\eeqn}{\end{eqnarray}}

\def\etal{ {\em et al.}~}

\def\q2 {q^2}

\begin{document}
\begin{titlepage}

\begin{flushleft}
\end{flushleft}
\begin{flushright}
{\large 
OITS-740
}
\end{flushright}

\begin{center}
{\Large\bf Higgs production in association 
with top quark pair at $e^+e^-$ colliders in theories of 
higher dimensional gravity}\\[10mm]
{\large Debajyoti Choudhury$^{a,b,}$\footnote{email: debchou@iacs.res.in, debchou@mri.ernet.in}
, N.~G.~Deshpande$^{c,}${\footnote{email: desh@oregon.uoregon.edu}}
and  
Dilip Kumar Ghosh$^{c,}${\footnote{email: dghosh@physics.uoregon.edu}}} 
\\[4mm]
{\em $^a$ Department of Theoretical Physics,
          Indian Association for the Cultivation of Science,\\
          2A\&B, Raja Subodh C. Mullick Road, Jadavpur, 
          Calcutta 700 032, India}\\[2mm]
{\em $^b$ Harish-Chandra Research Institute,\\
Chhatnag Road, Jhusi,
Allahabad 211019, India}\\[2mm]
{\em $^c$ Institute of Theoretical Science \\
     5203 University of Oregon\\
Eugene OR 97403-5203, U.S.A.} \\[10mm]
\end{center}
\begin{abstract}
The models of large extra compact dimensions, as suggested by Arkani-Hamed,
Dimopoulos and Dvali, predict exciting phenomenological consequences with 
gravitational interactions becoming strong at the TeV scale. Such theories
can be tested at the existing and future colliders. In this paper, we study 
the contribution of virtual Kaluza-Klein excitations  
in the process $e^+e^- \to t \bar t H $ at future linear collider (NLC). 
We find that the virtual exchange KK 
gravitons can modify the cross-section $\sigma(e^+e^- \to t \bar t H )$ 
significantly from its Standard 
Model value and will allow the effective string scale to be probed up to 
7.9 TeV. 
 \end{abstract}
 
\end{titlepage}

\vskip 1 true cm

\newpage

\def\baselinestretch{1.8}

\section{Introduction}
\label{sect:intro}

The concept of large extra dimensions and TeV scale gravity introduced
by Arkani-Hamed, Dimopoulos and Dvali, often referred to as the ADD
model~\cite{ADD} has attracted a lot of attention. In this scenario,
the total space-time has $D= 4+n $ dimensions, with the additional $n$
dimensions compactified. While gravity lives in the entire bulk
space-time, the Standard Model (SM) particles are deemed to be
confined to the usual four (i.e. $3+1$) dimensions. The 4-dimensional Planck
scale, $M_{Pl}\sim 2.4\times 10^{18}$~GeV , is no longer a fundamental
quantity but is derived from the size $R$ of the extra
dimensions\footnote{In principle, each of the extra dimensions could have 
  a different size. For the sake of simplicity though, we shall assume
  that the radii of the compactified $T^n$ are identical.}
and the {\it fundamental} Planck scale $M_{S}$ in the full theory:
\beqn
M^2_{Pl}\sim R^{n} M^{n +2}_{S} \ .
\eeqn 

Thus, for large extra-dimensions, it is possible to have a {\it
fundamental} scale $M_{S}$ as low as a TeV~\cite{ADD}, thereby
``solving'' the gauge hierarchy problem of the standard model.
However, $M_S \sim {\cal O} (1 \tev)$ in a model with a single extra
dimension ($n= 1$) necessitates $R \sim 10^{11} mm$, a value that
obviously runs counter to astrophysical observations. On the other
hand, for $n\geq 2 $, we have $R\leq $ mm, a range that is still
allowed by gravitational experiments.

The graviton couples to the standard model matter and gauge particles
through the energy-momentum tensor, with a strength suppressed by
powers of the $4$-dimensional Planck scale, $M_{Pl}$. However, from
the 4-dimensional point of view, the massless graviton propagating in
the $(4+n)$-dimensional bulk is to be interpreted as a  tower of
massive Kaluza-Klein (KK) modes of excitations, with spin-2, spin-1 (which
decouples ) as well spin-0 components.  In the context of collider
experiments, the mass spectrum of these KK modes can be treated as a
continuum, as the mass splitting ($\sim 1/R$) between successive modes
is about $10^{-4}$ eV (1 MeV) for $n=2 (6)$. On summing over the KK
modes, the effective graviton-exchange contribution to processes
involving the standard model particles is only suppressed by powers
of $(\sqrt{s}/M_S)$~\cite{ADD}, where $\sqrt{s}$ is the energy
available for the process. The Feynman rules for this theory may be
developed from a linearized theory of gravity in the bulk and may be
found in Refs.~\cite{GRW, Han-Lykken}.  These new interactions can
give rise to several interesting phenomenological consequences
testable at present and future colliders~\cite{ADD_rev} with their
 effects  observed either through production of
real KK modes, or through the exchange of virtual KK modes in various
processes~\cite{ADD_rev}.

The next generation $e^+e^-$ linear colliders 
is expected to function from 300 GeV up to about 1 TeV ( JLC, NLC, TESLA),
referred to as the LC \cite{JLC,NLC, TESLA}. There is also a possibility of
multi-TeV linear collider operating between energy range of 3-5 TeV at CERN
\cite{CLIC}. In this paper we will consider the process 
$e^+e^- \to t \bar t H $ to
study the effect of low-scale gravity at the proposed linear colliders
operating with a center of mass energy 500 GeV and beyond.  Within the
SM, $t \bar t H $ production has been studied in the context of the
determination of the top quark Yukawa couplings ~\cite{tth_born, top_yukawa}.
Thus, one would be looking for significant deviations from the
standard model expectations as a signal of new physics.  However, it
should be borne in mind that these processes  are unlikely to
serve as the dominant discovery channel for KK gravitons, since there
exist other simpler channels that are equally (or more) sensitive to
such graviton exchanges\cite{ADD_rev}. On the other hand, once a discovery is
made, the next phase would comprise of confirmatory tests as well as 
the determination of the parameters of the theory. This can be achieved only 
through a series of other experiments and this
is where the process under discussion would be useful.

\section{$e^+e^- \to t \bar t H $ }
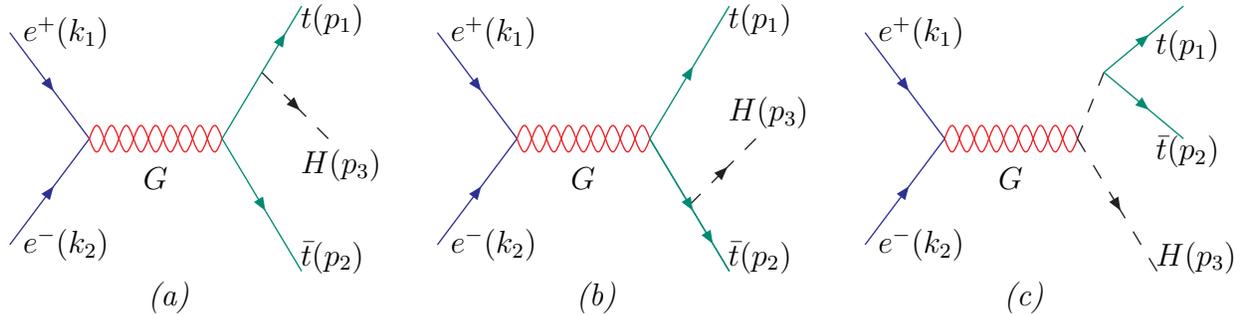
\begin{figure}[hbt]

\vspace*{-11ex}
 \begin{tabbing}
 \hspace*{-1.3em}
\begin{picture}(155,120)(-5.0,-20)
\ArrowLine(10, 40)(40,0){psBlue}
      \Text(15,40)[l]{$e^+(k_1)$}		
\ArrowLine(10, -40)(40,0){psBlue}
      \Text(15, -40)[l]{$e^-(k_2)$}		
\Photon(40,0)(90,0){5}{4.5}{psRed}
\Photon(90,0)(40,0){5}{4.5}{psRed}
      \Text(65,-15)[c]{$G$}		
\Line(90,0)(105,25){psPineGreen}
\ArrowLine(105,25)(120,50){psPineGreen}
      \Text(120,45)[l]{$t(p_1)$}		
\DashArrowLine(105,25)(130,0){5}{psBlack}
      \Text(120,-10)[l]{$H(p_3)$} 
\ArrowLine(90,0)(120,-50){psPineGreen}
      \Text(120,-45)[l]{$\bar t(p_2)$}
\Text(70, -60)[c]{\em (a)}
\end{picture}

\hskip 0.1cm

\begin{picture}(155,120)(-5.0,-20)
\ArrowLine(10, 40)(40,0){psBlue}
      \Text(15,40)[l]{$e^+(k_1)$}		
\ArrowLine(10, -40)(40,0){psBlue}
      \Text(15, -40)[l]{$e^-(k_2)$}		
\Photon(40,0)(90,0){5}{4.5}{psRed}
\Photon(90,0)(40,0){5}{4.5}{psRed}
      \Text(65,-15)[c]{$G$}		
\ArrowLine(90,0)(120,50){psPineGreen}
      \Text(120,45)[l]{$t(p_1)$}		
\Line(90,0)(105,-25){psPineGreen}
\ArrowLine(105,-25)(120,-50){psPineGreen}
\DashArrowLine(105,-25)(130,0){5}{psBlack}
      \Text(120,10)[l]{$H(p_3)$} 
\ArrowLine(90,0)(120,-50){psPineGreen}
      \Text(120,-45)[l]{$\bar t(p_2)$}
\Text(70, -60)[c]{\em (b)}
\end{picture}

\hskip 0.1cm

\begin{picture}(155,120)(-5.0,-20)
\ArrowLine(10, 40)(40,0){psBlue}
      \Text(15,40)[l]{$e^+(k_1)$}		
\ArrowLine(10, -40)(40,0){psBlue}
      \Text(15, -40)[l]{$e^-(k_2)$}		
\Photon(40,0)(90,0){5}{4.5}{psRed}
\Photon(90,0)(40,0){5}{4.5}{psRed}
      \Text(65,-15)[c]{$G$}		
\DashArrowLine(90,0)(120,-50){5}{psBlack}
      \Text(120,-45)[l]{$H(p_3)$} 
\DashLine(90,0)(100,25){5}{psBlack}
\ArrowLine(100,25)(130,50){psPineGreen}
      \Text(120,35)[l]{$t(p_1)$}
\ArrowLine(100,25)(130,0){psPineGreen}
      \Text(120,-5)[l]{$\bar t(p_2)$}
\Text(70, -60)[c]{\em (c)}
\end{picture}
\end{tabbing}

\vspace*{3ex}
   \caption{\em The graviton-mediated diagrams contributing to 
    $e^+ e^- \rightarrow t \bar t H$ at the tree level.}
 \label{Fig:feyn}
\end{figure}

In this section we study the effect of the graviton exchange in the 
production of a Higgs particle in association with a pair of 
top quarks at a Linear Collider. The SM diagrams contributing to this
process are well known. In Fig.\ref{Fig:feyn}, we present the additional 
set of diagrams that arise in the theory under consideration. 
Note that, in principle, there exists yet another diagram involving 
the 4-point vertex $t \bar t H G$. However, only the trace of the graviton 
appears in this diagram, and in the limit of vanishing electron mass 
(an excellent approximation), the contribution disappears identically.

The amplitudes corresponding to the diagrams in Fig.\ref{Fig:feyn} can
be easily calculated following the Feynman rules derived in,
say, Ref.\cite{Han-Lykken}.  As mentioned earlier, the gravitational
coupling $\kappa \equiv \sqrt{16 \pi G_N}$ ($G_N$ being the
four-dimensional Newton's constant) can be expressed in terms of the
fundamental scale $M_S$ and the size $R$ of the $n$ extra dimensions
through
\bea
\kappa^2 R^n = 16 \pi (4\pi)^{n/2} \Gamma(n/2) M_S^{-(n+2)}
\eea
Choosing to work in the de Donder gauge, we have, for the amplitudes, 
\bea
{\cal M}_{a}&=&  \frac{-{\cal G}g m_t \pi}{16 M_W} \:
          \frac{P^{\mu\nu\alpha\beta}}{(p_1 + p_3)^2 - m_t^2} \:
      \big [{\bar v(k_1)}C_{\mu\nu} u(k_2) \big]  \:
      \big[{\bar u(p_1)} \:
           (p_1{\!\!\!\!/}+ p_3{\!\!\!\!/} + m_t ) 
             {\chi}^{a}_{\alpha\beta} v(p_2) \big] \nonumber \\
{\cal M}_{b}&=&  \frac{{\cal G}g m_t \pi}{16 M_W} \:
          \frac{P^{\mu\nu\alpha\beta}}{(p_2 + p_3)^2 - m_t^2} \:
      \big [{\bar v(k_1)}C_{\mu\nu} u(k_2) \big] \:
      \big[{\bar u(p_1)} \:
           (p_2{\!\!\!\!/}+p_3{\!\!\!\!/} - m_t ) 
               {\chi}^{b}_{\alpha\beta} v(p_2) \big]\\
{\cal M}_{c}&=& \frac{-{\cal G} g m_t \pi}{ 2 M_W} \:
         \frac{P^{\mu\nu\alpha\beta}}{(p_1+p_2)^2 - m_H^2} \:
      \big [{\bar v(k_1)}C_{\mu\nu} u(k_2) \big] \:
      \big[{\bar u(p_1)}{\chi}^{c}_{\alpha\beta}
 v(p_2) \big]    \nonumber
    \label{mat_elem}
\eea
where, 
\beqn
\begin{array}{rcl}
{\cal G} & =& \dis \frac{\kappa^2}{16 \pi} D(s)  
         =  M_S^{-4} \: \left( \frac{\sqrt{s}}{M_S} \right)^{n-2} 
             \: \left[ \pi + 2 i \: P \int_0^{M_S / \sqrt{s}} 
                                       \frac{y^{n -1} \: dy}{1 - y^2}
                \right]
\\[2ex]
P^{\mu\nu\alpha\beta} & = & \dis
     \eta^{\mu\alpha}\eta^{\nu\beta} + \eta^{\mu\beta}\eta^{\nu \alpha} 
- \frac{2}{3}\eta^{\mu\nu}\eta^{\alpha\beta}   \\
C_{\mu\nu} & = & \dis
    \big[\gamma_\mu \big(k_2 - k_1)_{\nu} + (\mu \leftrightarrow \nu) \big] \\
{\chi}^{a}_{\alpha\beta} & = & \dis
     \left[ \gamma_{\alpha} \big(p_1-p_2+p_3)_{\beta}
            - \eta_{\alpha\beta} 
           \big( p_1{\!\!\!\!/}-p_2{\!\!\!\!/}+p_3{\!\!\!\!/}- 2 m_t\big)
     \right]
       + (\beta \leftrightarrow \alpha ) \\ 
{\chi}^{b}_{\alpha\beta} & = & \dis
       \left[ \gamma_{\alpha} \big(p_1-p_2-p_3)_{\beta}
              - \eta_{\alpha\beta} 
               \big( p_1{\!\!\!\!/}-p_2{\!\!\!\!/}-p_3{\!\!\!\!/}- 2 m_t\big) 
                     \right]
       + (\beta \leftrightarrow \alpha ) \\ 
{\chi}^{c}_{\alpha\beta} &= & m_H^2 \eta_{\alpha\beta} - 
(p_1+p_2)^\mu p_3^\nu \left[ \eta_{\mu \alpha} \eta_{\nu \beta} 
                             + \eta_{\nu \alpha} \eta_{\mu \beta}
                             - \eta_{\mu \nu} \eta_{\alpha \beta} \right]
\end{array}
    \label{mat_defns}
\eeqn   
with $s \equiv (k_1 + k_2)^2 $. The function $D(s)$ is the resultant of summing
over the KK modes of the graviton, and only the principal 
part of the integral is to be taken.

\begin{figure}[hbt]
\vspace*{-0.7cm}
\centerline{
\epsfxsize = 22cm  \epsfysize = 8cm \epsfbox{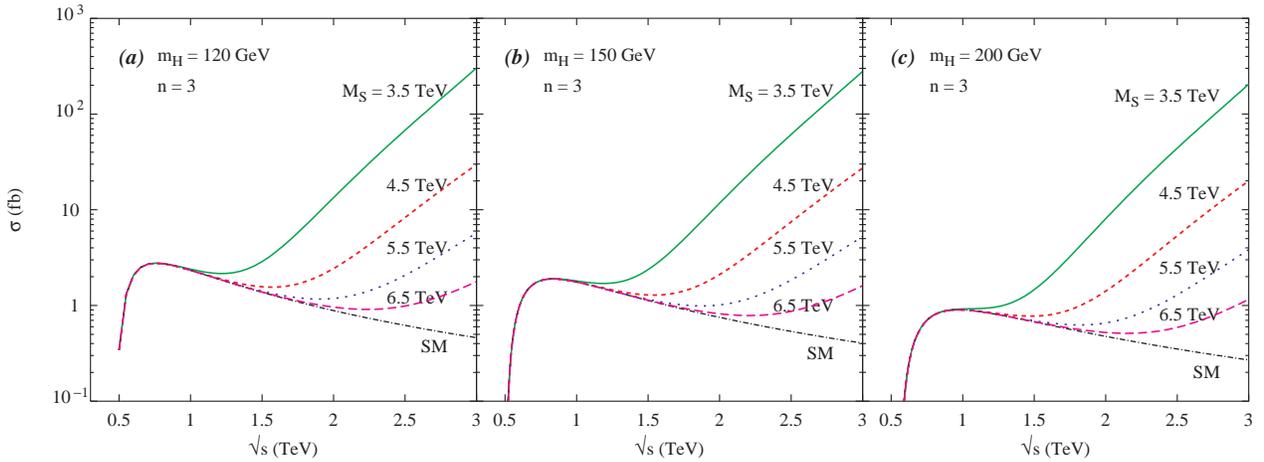}
}

\vspace*{-1.2cm}
\caption{\em Variation of the cross-section (fb) for  
$t \bar t H$ production with 
machine energy. The four dashed curves correspond to the ADD predictions
for $M_S = 3.5, 4.5$ , $5.5 $ and $6.5$ TeV and $n=3$, while the solid line 
represents the standard model expectations. The three panels correspond
to three different values of the Higgs mass.}
  \label{Fig:cs_ener}
\end{figure}

The cross sections after adding the new amplitudes to the SM 
are presented in Fig.\ref{Fig:cs_ener} as
a function of machine energy, and for different Higgs masses. 
To be specific, we hold the number of
extra dimensions to be $n = 3$.  In each figure, the solid line
represents the standard model predictions and exhibits the expected
fall-off with energy. The inclusion of the graviton exchange changes
the $\sqrt{s}$-dependence drastically.  For relatively large values of
$M_S$, the fall-off persists upto a certain value of $\sqrt{s}$ and
thereafter the cross section increases rapidly. And, as expected, the
turnaround point (in $\sqrt{s}$) is a monotonic function of
$M_S$. Thus we can expect stronger bounds on $M_S$ as the machine
energy increases.  Comparing the three panels in Fig.\ref{Fig:cs_ener}, 
it is very
clear that the ADD contribution is not strongly dependent on Higgs
mass. At low energies, the difference in the magnitude of
cross-sections for these three values of Higgs masses arises mainly
from the phase space. At relatively higher value of $\sqrt{s}$, the
cross-sections corresponding to $m_H = 120$ and 200 GeV become nearly
identical.  Indeed, the relative difference between the cross sections
for any two different values of $m_H$ tends to be smaller in the ADD
case than that within the standard model.

 \begin{figure}[hbt]
\vspace*{-2cm}
 \centerline{
\epsfxsize = 8cm  \epsfysize = 10cm \epsfbox{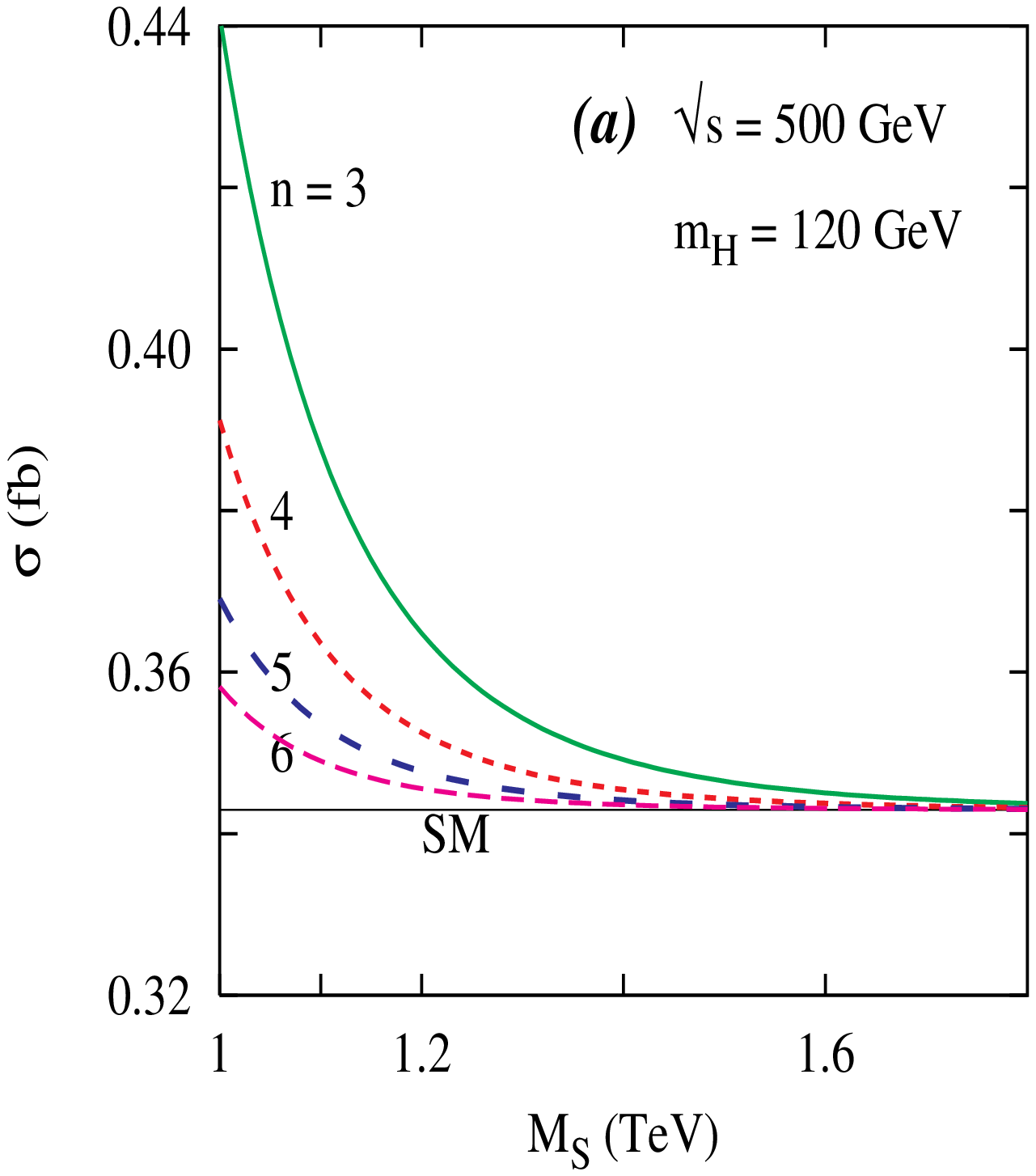}
\epsfxsize = 8cm  \epsfysize = 10cm \epsfbox{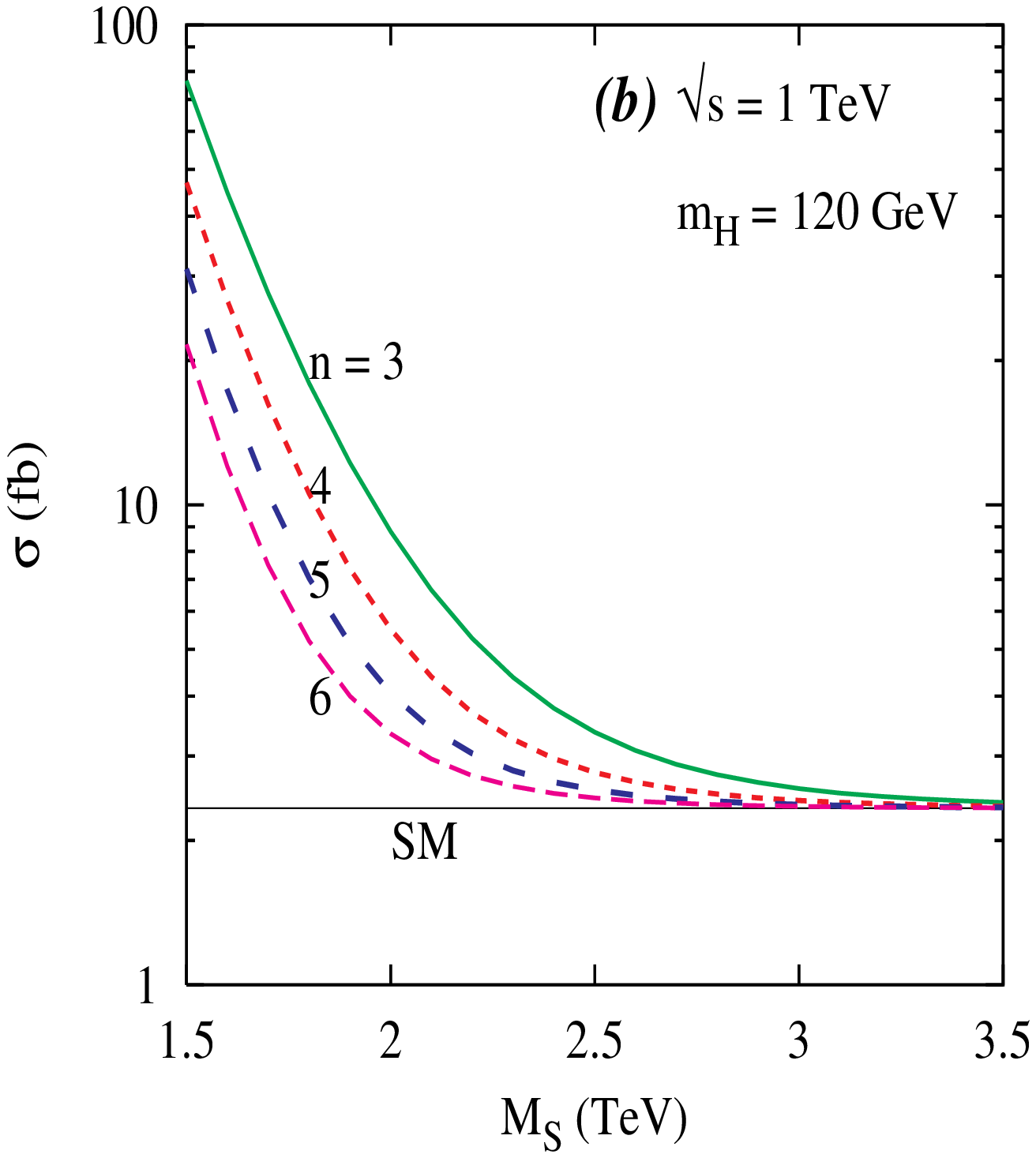}
}
 \vspace*{-0.2cm}
 \caption{\em The variation of the cross-section (fb) for  
 $t \bar t H$ production with $M_S$, for a Higgs mass of 120 GeV.
 The curves correspond to the ADD predictions
for $n = 3, 4, 5 $ and $6$ respectively. Also indicated are
the standard model expectations.
The left and the right panels correspond to $\sqrt{s}= 500$ GeV and 1 TeV
respectively.}
\label{Fig:cs_MS}
\end{figure}

Since the integral in $D(s)$---see eq.(\ref{mat_defns})---is a
relatively slowly increasing function of $(M_S / \sqrt{s})$, the
amplitudes typically go down as $\sim M_S^{-4}$. This is reflected in
Fig.\ref{Fig:cs_MS}, where we present the total cross-section as a
function of $M_S$ for theories with different numbers of extra
dimensions.  For comparison, the SM cross-sections are also given.
As with simpler processes, here too 
the graviton contribution decreases with the increase in
the number of extra-dimension as long as the machine energy and
$M_S$ are held constant.

It is thus obvious that graviton-exchange contributions can 
significantly alter the predictions for  the $e^+ e^- \to t\bar t H$ 
cross-section in such models. However, before one can claim that any 
such observed deviation has been occasioned by the exchange of virtual 
gravitons, one needs to ascertain that such deviations cannot be explained 
by any other possible 4-dimensional new physics effects (4DNP) going 
beyond the SM. We proceed to do this next.

Note that, within the SM, the process under discussion is controlled by 
two relatively undetermined couplings, namely the $t \bar t H$ and the
$ Z Z H$ ones. Of these, the latter is expected to be measured to 
a very great accuracy at such a collider~\cite{zzh_coupling}. As for 
the former, the expected accuracy is far less. In fact, the process under
consideration, in itself, is perhaps the best channel for this measurement!
In view of this ignorance, let us assume that any 4DNP effect may 
change the $t\bar t H$ coupling by at most $10 \%$. While 
this may be an underestimate, we shall see that even a somewhat larger
uncertainty would not change our conclusions qualitatively. 
 \begin{figure}[!h]
\vspace*{2ex}
 \centerline{
\epsfxsize = 8cm  \epsfysize = 10cm \epsfbox{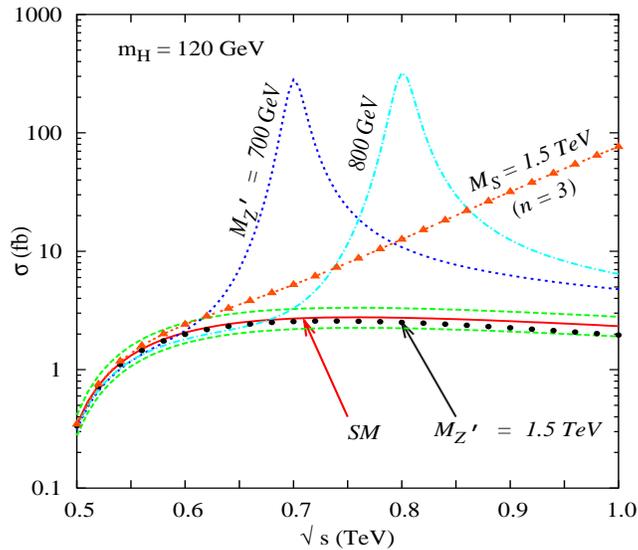}
}
 \vspace*{-3.5cm}
 \caption{\em Variation of $\sigma(e^+e^-\to t \bar t H)$ with machine 
energy ($m_H = 120 \ GeV$).
 The solid curve represents the SM cross-section while the band around it
indicates the variation in it wrought by a 
$\pm 10\%$ deviation in the SM $t\bar t H$ coupling. The monotonically
increasing curve corresponds to a 
ADD model with  $n = 3$ and $M_S = 1.5$ TeV. Also shown are three curves
corresponding to a 4-dimensional 
model with a sequential $Z'$ for different values 
of the latter's mass.}
\label{Fig:tth_dev}
\end{figure}

In Fig.\ref{Fig:tth_dev}, we display the effect of such a variation in the 
$t\bar t H$ coupling on the cross-section under consideration.
The two sidebands around the central (SM) curve represent the 
cross section expected for an enhancement (reduction) of the 
$t\bar t H$ coupling by $10\%$. 
As is immediately obvious, 
the functional dependence of the cross-section on the
center-of-mass energy remains virtually  the same as in the standard model 
for modest changes in this coupling.
On the other hand, the functional dependence is
markedly different for ADD scenarios. Thus, measurements at two
different center of mass energies would clearly distinguish 
between these two classes of models. 

As a further example of 4DNP, let us consider an extension of the
SM to a model with an extra neutral gauge boson ($Z'$). For illustrative 
purposes, we choose to work with a sequential $Z'$, namely one whose
couplings to the SM particles are exactly the 
same\footnote{The corresponding results for a $Z'$ with a different 
  set of couplings are qualitatively very similar.}
 as those of the 
usual $Z$ \cite{leike}. In Fig.\ref{Fig:tth_dev}, we also demonstrate how the $t \bar t H$
production cross section may change in the presence of such a $Z'$. 
For center of mass energies well below $m_{Z'}$, the cross section would
naturally be quite similar to that within the SM. However, as the energy 
is raised, the cross section starts to grow rapidly, potentially mimicking
the effect within an ADD model. However, if the energy is raised further,
the resonance structure characteristic of such a $Z'$ would evince itself 
(as also in other production modes). If the energy were to be raised still 
further to values much larger than $m_{Z'}$, the cross section would fall
as dictated by considerations of partial wave unitarity. Thus, once again,
such models are easily distinguishable from ADD-like ones.

Now that a method of distinguishing has been delineated,
we may try to estimate the sensitivity of the process under 
consideration. To this end, we display, in Table 1,
the $3\sigma$ exclusion limit on $M_S$ for different machine 
energies and number of extra dimensions.
 In estimating these limits we assume an
integrated luminosity of $500~{\rm fb}^{-1}$, 
detection efficiency of $80\%$ and a $2\%$ systematic error.
We would, however, like to point out that, in the absence of a 
detailed simulation, such estimates are necessarily very crude ones. 
Furthermore, while estimating the SM $tth$ cross-sections we have not 
considered QCD and EW corrections, which has been computed by several
groups \cite{tth_strong, tth_ew}.  

\begin{table}
\begin{center}
\begin{tabular}{|c|c|c|c|c|}
\hline
$\sqrt{s}$&\multicolumn{4}{|c|}{$M_S$ in GeV}\\[3mm]
\cline{2-5}
\vspace*{-2ex}TeV&$n=3$&4&5&6\\[3mm]
\hline
&&&&\\
0.5  & 920 (880) & 855 (825) & 810 (785) & 780 (750) \\[3mm]
1.0& 2530 (2470) & 2300 (2240) & 2140 (2090) & 2020 (1970) \\[3mm]
3.0& 7970 (7680) & 7200 (6960) & 6680 (6470) & 6300 (6100) \\[3mm]
\hline
\end{tabular}
\caption{\em $3\sigma $ exclusion limits on $M_S$ achievable 
with an integrated luminosity
of $500~{\rm fb}^{-1}$ for different center-of-mass energies 
($\sqrt{s}$). $n$ is the number of extra dimensions and we assume 
$m_H = 120 $ GeV. The numbers in parentheses refer to the corresponding
($5 \sigma$) discovery limits.
}
\end{center}
\label{exclusion}
\end{table}

That the exclusion limits vary strongly with the available 
energy is but a consequence of the structure of the theory (in particular,
the momentum dependence of the couplings) and could have already been
expected from Figs.~\ref{Fig:cs_ener} or ~\ref{Fig:cs_MS}. What is 
also remarkable is that the exclusion (discovery) limits on $M_S$ 
are only weakly dependent on the number of extra dimensions.

\section{Summary}

In this paper we have studied the implications of KK graviton contribution
to the process $e^+e^- \to t \bar t H$, which has been studied in the standard 
model. The spin-2 mode of KK gravitons contribute to this
process substantially. It is likely though that 
the existence of low-energy quantum 
gravity may be discovered through more direct channels as studied by
several groups. Nevertheless, this process will be an independent 
confirmation for such a discovery. We have shown both $3\sigma $ exclusion 
limit and  $5\sigma$ discovery limit for the string scale $M_S $ 
obtainable at typical linear collider energies assuming 
the benchmark integrated luminosity of $500~{\rm fb}^{-1}$.

\section*{Acknowledgments}
The authors thank the High Energy Physics Division of the Argonne
National Laboratory for hospitality during the period the project was
initiated.  DC thanks the Department of Science \& Technology, India
for financial assistance under the Swarnajayanti Fellowship grant.
The work of NGD and DKG was supported by US DOE contract numbers
DE-FG03-96ER40969.

\def\pr#1,#2 #3 { {Phys.~Rev.}        ~{\bf #1},  #2 (19#3) }
\def\prd#1,#2 #3{ { Phys.~Rev.}       ~{D \bf #1}, #2 (19#3) }
\def\pprd#1,#2 #3{ { Phys.~Rev.}      ~{D \bf #1}, #2 (20#3) }
\def\prl#1,#2 #3{ { Phys.~Rev.~Lett.}  ~{\bf #1},  #2 (19#3) }
\def\pprl#1,#2 #3{ {Phys. Rev. Lett.}   {\bf #1},  #2 (20#3)}
\def\plb#1,#2 #3{ { Phys.~Lett.}       ~{\bf B#1}, #2 (19#3) }
\def\pplb#1,#2 #3{ {Phys. Lett.}        {\bf B#1}, #2 (20#3)}
\def\npb#1,#2 #3{ { Nucl.~Phys.}       ~{\bf B#1}, #2 (19#3) }
\def\pnpb#1,#2 #3{ {Nucl. Phys.}        {\bf B#1}, #2 (20#3)}
\def\prp#1,#2 #3{ { Phys.~Rep.}       ~{\bf #1},  #2 (19#3) }
\def\zpc#1,#2 #3{ { Z.~Phys.}          ~{\bf C#1}, #2 (19#3) }
\def\epj#1,#2 #3{ { Eur.~Phys.~J.}     ~{\bf C#1}, #2 (19#3) }
\def\mpl#1,#2 #3{ { Mod.~Phys.~Lett.}  ~{\bf A#1}, #2 (19#3) }
\def\ijmp#1,#2 #3{{ Int.~J.~Mod.~Phys.}~{\bf A#1}, #2 (19#3) }
\def\ptp#1,#2 #3{ { Prog.~Theor.~Phys.}~{\bf #1},  #2 (19#3) }
\def\jhep#1, #2 #3{ {J. High Energy Phys.} {\bf #1}, #2 (19#3)}
\def\pjhep#1, #2 #3{ {J. High Energy Phys.} {\bf #1}, #2 (20#3)}

\end{document}